\documentclass[12pt,titlepage]{article}
\pdfoutput=1

\usepackage[utf8]{inputenc} 
\usepackage[T1]{fontenc} 
\usepackage{lmodern} 

\usepackage{setspace} 
\usepackage{amsmath}

\usepackage[pdftex]{graphicx}
\usepackage{color}
\usepackage[normalem]{ulem} 
\usepackage[round,numbers,sort&compress]{natbib} 
\usepackage{textcomp} 
\usepackage{pdfpages}
\usepackage{epsfig}
\usepackage{tabularx}
\usepackage{booktabs}
\usepackage{multirow}
\usepackage{siunitx} 
\usepackage[version=3]{mhchem} 
\usepackage{color}
\usepackage{pdfpages}

\DeclareSIUnit{\weightpercent}{wt\%}
\DeclareSIUnit{\molpercent}{mol\%}
\DeclareSIUnit{\zJ}{\zepto\J}

\newcommand{\changed}[1]{{#1}} 

\begin{document}

\noindent\LARGE{\textbf{Bending rigidities and interdomain forces in membranes with coexisting lipid domains}}
\vspace{0.6cm}

\noindent\large{Benjamin Kollmitzer,$^{\dagger, \ddagger}$ Peter Heftberger,$^{\dagger, \ddagger}$ Rudolf Podgornik$^{\S,\P,\|}$, John F. Nagle$^{**}$, and Georg Pabst$^{\dagger, \ddagger}$ $^{\ast}$}\vspace{0.5cm}

\noindent{\small{\textit{$^{\dagger}$~	University of Graz, Institute of Molecular Biosciences, Biophysics Division, NAWI Graz, Humboldstr. 50/III,	A-8010 Graz, Austria.}

\noindent\textit{$^{\ddagger}$~ BioTechMed-Graz, Austria.}

\noindent\textit{$^{\S}$~ Department of Theoretical Physics, Jozef Stefan Institute, Ljubljana, Slovenia.}

\noindent\textit{$^{\P}$~ Department of Physics, Faculty of Mathematics and Physics, University of Ljubljana, Ljubljana, Slovenia.}

\noindent\textit{$^{\|}$~ Department of Physics, University of Massachusetts, Amherst, Massachusetts, USA.}

\noindent\textit{$^{**}$~ Department of Physics, Carnegie Mellon University, Pittsburgh, Pennsylvania, USA.}

\vspace{2 cm}

\noindent{\footnotesize $^{\ast}$~Correspondence: georg.pabst@uni-graz.at}

\abstract{In order to precisely quantify the fundamental interactions between heterogeneous lipid membranes with coexisting liquid-ordered (Lo) and liquid-disordered (Ld) domains, we performed detailed osmotic stress SAXS experiments by exploiting the domain alignment in raft-mimicking lipid multibilayers. Performing a Monte Carlo (MC) based analysis allowed us to determine with high reliability the magnitude and functional dependence of interdomain forces concurrently with the bending elasticity moduli. In contrast to previous methodologies, this approach enabled us to consider the entropic undulation repulsions on a fundamental level, without having to take recourse to crudely justified mean-field like additivity assumptions. Our detailed Hamaker coefficient calculations indicated only small differences in the van der Waals attractions of coexisting Lo  and Ld  phases. In contrast, the repulsive hydration and undulation interactions differed significantly, with the latter dominating the overall repulsions in the Ld phase. \changed{Therefore, alignment of like domains in multibilayers appears to originate from both, hydration and undulation repulsions.}}


\clearpage



\section{Introduction}
Diverse physiological processes in living systems depend on fundamental physical interactions between lipid membranes acting on the nanoscopic length scale.
Of particular interest in this context \changed{are, besides intra-membrane interactions \cite{semrau_membrane-mediated_2009, ursell_morphology_2009},} forces acting between membrane domains/rafts across the aqueous phase, \changed{which are also involved} in their correlated mutual alignment. 
Such positional correlations are well established for liquid-ordered (Lo)/liquid-disordered (Ld) domains in model lipid multibilayers \cite{tayebi_long-range_2012, karmakar_structure_2005, chen_partition_2007, mills_liquid-liquid_2008, staneva_role_2008, pabst_effect_2009, yuan_solution_2009, uppamoochikkal_orientation_2010, heftberger_situ_2015}\changed{. 
Several groups have established compositional phase diagrams for mixtures of high-melting lipid, low-melting lipid and cholesterol, which exhibit Lo/Ld phase coexistence over a broad range of compositions and temperatures \cite{heberle_phase_2011, marsh_cholesterol-induced_2009}. These systems mimic mammalian outer plasma membranes and enable studies of domain properties under well-defined conditions. Most recently, we reported structural details of Lo/Ld phases in two ternary lipid mixtures using a global small-angle X-ray scattering (SAXS) analysis for coexisting lipid domains \cite{heftberger_situ_2015}. This analysis relies on the above mentioned mutual alignment of like domains. Domain-alignment is, however, also of biological relevance, for example in the context of}, 
the immune response, where organization of receptor--ligand domains occurs during T-cell adhesion \cite{monks_three-dimensional_1998, grakoui_immunological_1999}. Both, the formation of such domains as well as the adhesion affinity depend strongly on thermal fluctuations and consequently on the bending rigidity of membranes \cite{rozycki_segregation_2010, hu_binding_2013}. It is therefore reasonable to expect that fundamental intermembrane interactions will play an important role also in receptor--ligand domain alignment. 

Within the broad DLVO paradigm \cite{israelachvili_interactions_2011} the fundamental long-range interactions between soft material interfaces, mediated by their molecular environment, such as solvation (hydration) interaction, electrostatic interaction, and van der Waals interaction, can be treated independently and additively. However, this additivity {\it Ansatz} is in general not vindicated for entropically driven bending undulation interactions that warrant a more sophisticated approach \cite{helfrich_steric_1978, podgornik_thermal-mechanical_1992, israelachvili_interactions_2011}.

Besides the fundamental role of entropic membrane undulations, their relation with the membrane bending rigidity $K_c$ \cite{helfrich_steric_1978}, and through it their connection with diverse physiological processes, has spurred a sustained scientific interest \cite{pabst_coupling_2013}. Shape analysis of giant unilamellar vesicles (GUV) \cite{meleard_bending_1997}, diffuse X-ray scattering from oriented lipid multibilayers \cite{lyatskaya_method_2000}, and GUV micropipette aspiration \cite{evans_entropy-driven_1990} are all techniques exploiting this connection, but none of them so far has been able to simultaneously determine the bending moduli for coexisting membrane phases. On the other hand, macroscopically sized domains form distinct lamellar lattices in multibilayer systems, enabling the application of \emph{osmotic stress experiments} \cite{pabst_effect_2009, boulgaropoulos_lipid_2012}. In such experiments, osmotic pressure is maintained by, e.g., \ large neutral polymers, such as poly-ethylene-glycol (PEG), which do not penetrate into the interbilayer water layer, while the corresponding bilayer separation and more recently also the specific line broadening due to fluctuations are measured by small-angle X-ray scattering (SAXS). Several groups, including ours, have previously applied this approach to study interactions between macromolecules, including lipid bilayers \cite{leneveu_measurement_1977, parsegian_measured_1979, parsegian_osmotic_1986, mcintosh_hydration_1986, mcintosh_contributions_1993, rand_hydration_1989, petrache_interbilayer_1998, pabst_entropy-driven_2007, pabst_effect_2009, pabst_applications_2010, boulgaropoulos_lipid_2012}. 

The bare long-range DLVO interaction components, that couple macromolecular surfaces through their molecular environment, get inextricably intertwined through the thermally driven conformational fluctuations of the soft interfaces, making detailed predictions of the overall interaction nearly impossible. Therefore, many studies in the past have resorted to describe such complicated thermal fluctuation effects by different mean-field/additivity approximations, where conformational fluctuation effects on the bare interaction potentials are included self-consistently \cite{helfrich_steric_1978, sornette_importance_1986, evans_thermal-mechanical_1986, podgornik_thermal-mechanical_1992, mecke_fluctuating_2003}. 
In contrast, additivity/mean-field approximations can be altogether avoided in the case of simulations that start from fundamental long-range DLVO interaction components and need no additional approximations to yield an accurate estimate for the total osmotic pressure in the system \cite{gouliaev_simulations_1998, gouliaev_simulations_1998-1}. 

In order to understand the coupling between bare interactions and thermal undulations \changed{in phase separated systems}, we apply a gradient-based optimization algorithm to iteratively adjust the parameters entering MC simulations, i.e.,\ the coefficients describing the strength and range of intermembrane interactions as well as the bending rigidity characterizing the thermal undulations, in order to best match simulation results with the experimental osmotic stress data for \changed{coexisting Lo/Ld phases. 
We demonstrate the capability of the simulation-driven analysis choosing a well-studied mixture of  dioleoyl phosphatidylcholine (DOPC), distearoyl phosphatidylcholine (DSPC) and cholesterol (Chol) \cite{zhao_phase_2007, heberle_comparison_2010, heberle_bilayer_2013}, previously shown to exhibit Lo/Ld domain alignment in the phase coexistence regime \cite{heftberger_situ_2015}. We find that Lo domains are about three times more rigid than Ld domains, which exhibit significant contributions to domain repulsion from bending fluctuations. On the other hand, hydration forces decay much slower with domain separation between Lo domains. In turn, attractive van der Waals interactions were found to be of similar magnitude between Lo--Lo and between Ld--Ld domains. Our results provide insight into the strength and distance dependence of forces at play between like-domains as a prerequisite to devising theories for domain alignment.}



\section{Materials and methods}
\subsection{Sample preparation}
DSPC, DOPC, and Chol were purchased from Avanti Polar Lipids, Inc., Alabaster, AL, USA and used without further purification. Poly(ethylene glycol) (PEG) with an average MW of \num{8000} was obtained from Fluka Chemie AG, Buchs, Switzerland and used as received. 

After weighing, lipids were dissolved in chloroform\slash methanol 2:1 at concentrations of \SI{10}{\mg \per \ml} \cite{note__supplier_provided_M_W}. We prepared the ternary lipid-only mixture DOPC\slash DSPC\slash Chol (0.42:0.37:0.21) in a glass vial and evaporated the organic solvent under a gentle nitrogen stream at \SI{30}{\degreeCelsius}\changed{. This lipid composition and its tie-line lie well inside the Lo/Ld phase coexistence region according to \cite{zhao_phase_2007, heberle_comparison_2010}, and the domains' structural properties have already been investigated with different methods \cite{heberle_bilayer_2013, heftberger_situ_2015}.} Remaining solvent traces were removed by placing the samples in vacuum overnight. \SI{18}{\Mohm \per \cm} water (UHQ PS, USF Elga, Wycombe, UK) was added at \SI{20}{\ul~water \per \mg~lipid} and the mixtures fully hydrated at \SI{50}{\degreeCelsius} for \SI{4}{hours} with repeated freeze-thaw cycles. 

To exert osmotic pressure on MLVs, \changed{samples were cooled to room temperature after hydration and aliquots overlayed} with PEG dissolved in water, yielding final concentrations of \SIrange{1}{42}{\weightpercent} PEG in water. Samples were protected against oxidation with argon, the vials closed and taped, and stored at \SI{4}{\degreeCelsius} for \SIrange{7}{10}{days} until the measurement. The osmotic equation of state for PEG, connecting its osmotic pressure with its solution concentration is well known \cite{cohen_phenomenological_2009} and allows for an accurate determination of the PEG osmotic pressure $P$ by using previously reported high resolution data \cite{stanley_measuring_2003}. 

\subsection{X-ray measurements} \label{subsec:saxs-measurements}
Small-angle X-ray scattering (SAXS) was performed at the Austrian SAXS beamline at ELETTRA, Trieste, Italy \cite{amenitsch_first_1998, bernstorff_high-throughput_1998}, at a wavelength of \SI{1.54}{\angstrom} and an energy-dispersion $\Delta E/E$ of \num{2.5e-3}. A mar300 Image Plate 2D detector from marresearch, Norderstedt, Germany was used, covering a $q$-range from \SIrange{0.2}{7.1}{\angstrom^{-1}} and calibrated with silver-behenate (\ce{CH_3(CH_2)_{20}-COOAg}) with a $d$-spacing of \SI{5.838}{\nm} \cite{huang_x-ray_1993}. Samples were filled into reusable quartz-glass capillaries and kept in a brass sample holder connected to a circulating water bath from Huber, Offenburg, Germany. The samples were equilibrated for \SI{10}{\minute} at \SI{20.0(1)}{\degreeCelsius} before exposing them for \SI{30}{\s} to the X-ray beam. 

The two dimensional detector signal was radially integrated with FIT2D \cite{hammersley_fit2d:_1997, hammersley_two-dimensional_1996}. Water background subtraction for samples without PEG was performed with Primus \cite{konarev_primus:_2003}. For osmotically stressed samples however, additional scattering from PEG made a standard background subtraction impractical. Since the essential informations in this case were just the Bragg peaks' shapes and positions, we subtracted approximative backgrounds, obtained by interpolating between SAXS signals of pure water and PEG/water mixtures. Alternatively, one could just subtract an arbitrary smooth function from the measured data. 

The reduced data were then fitted using a recently published, full $q$-range analysis method for coexisting liquid\slash liquid membrane domains \cite{note__tieline_endpoint_checks}. This method models each phase's contribution individually with a bilayer-structure and a superimposed membrane lattice. The lattice description is based on a modified Caill\'{e} theory \cite{zhang_theory_1994, caille_physique_1972} and therefore yields the average membrane periodicity $d$ and the line shape parameter $\eta$, which is connected to the mean square fluctuation of the membrane spacing via $\Delta^2 = \eta d^2 / \pi^2$ \cite{petrache_interbilayer_1998}. The bilayer-structure of each phase is then modeled separately via probability distributions of quasi-molecular fragments \cite{kucerka_lipid_2008}. 

Most importantly, the full $q$-range analysis allowed us to quantify the magnitude of fluctuations for coexisting domains. For both phases of stress-free samples, this also yields accurate electron density profiles, from which the bilayer thickness could be obtained; but this was not possible when osmotic pressure was applied. Instead, the osmotic thickening of $d_B$ was calculated using $d_B(P) = d_B(0) \cdot (K_A + P \cdot d(P))/(K_A + P \cdot d_B(0))$ \cite{rand_hydration_1989}, where the area extension modulus $K_A$ was estimated from published micropipette aspiration experiments on single lipids and binary lipid mixtures \cite{rawicz_elasticity_2008, rawicz_effect_2000}, as detailed in Sec.~S1 of the Supporting Material. The overall analysis was rather insensitive to uncertainties in $K_A$ because the maximal change in bilayer thickness was only slightly larger than the uncertainty of the fit ($\pm 2 \%$). The definition of the bilayer thickness $d_B$ was found to be more important. In principle one could determine optimal values of $d_B$ via a joint fit with free MC parameters, but this problem is under-determined and led to bizarre values of $d_B$ for different data sets \cite{gouliaev_monte-carlo_1998}. Instead, we defined $d_B$ as the distance between the remotest lipid atoms \cite{note__steric_headgroup_definition}, also known as the steric bilayer thickness \cite{mcintosh_hydration_1986}; this yielded good fits and comparable results, while being directly accessible from the SAXS analysis.

\subsection{Membrane Monte-Carlo simulation} \label{subsec:membrane_mc}
The simulation code used has been described previously in detail for a single membrane between two walls and for a stack of membranes \cite{gouliaev_simulations_1998, gouliaev_simulations_1998-1, gouliaev_monte-carlo_1998}. For completeness, but also to highlight our modifications, we briefly summarize its basic elements. 

\begin{figure*}
  \includegraphics[width=1.0\textwidth]{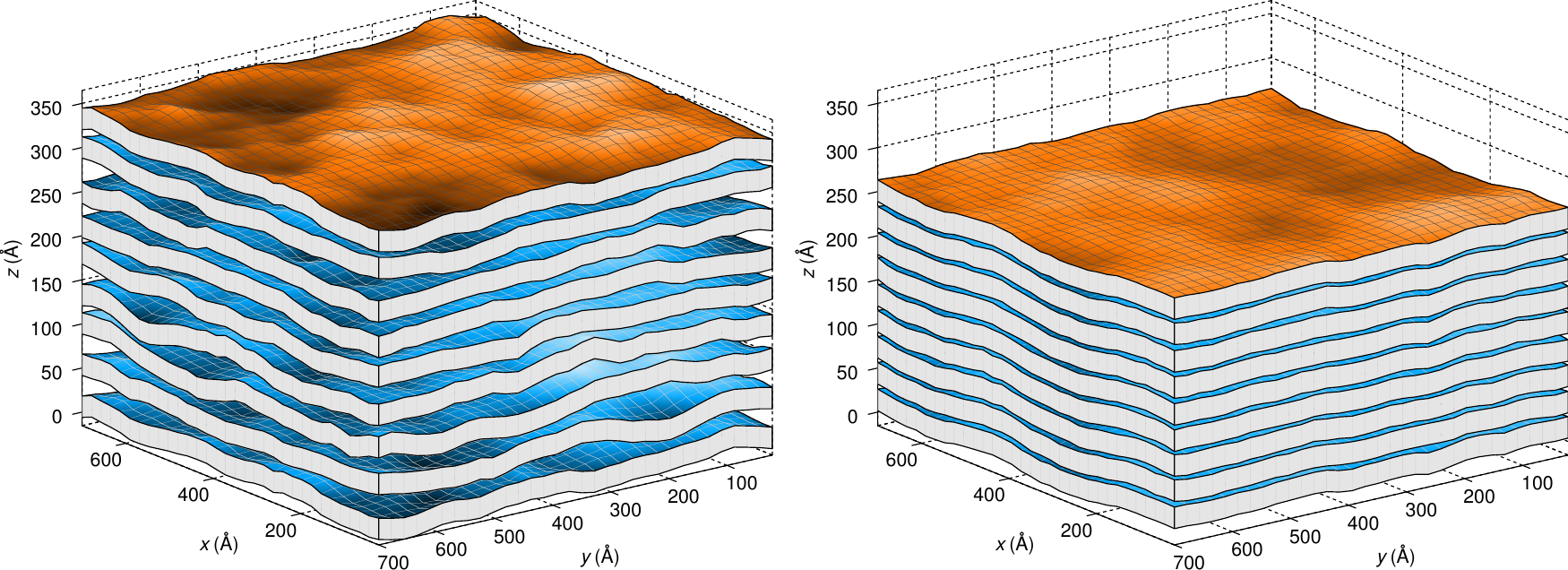}
  \caption{Real space snapshots of equilibrated Ld simulations at zero (left) and finite (\SI{5.5}{\MPa}, right) osmotic pressure. Membranes are drawn with their average thickness. Deviations from the periodic lattice are color coded. Due to 3D periodic boundary conditions, the top-most (orange) and bottom-most membranes are equal. The most prominent effects of external pressure, a compression of the stack and a reduction of the fluctuations, are clearly visible.}
  \label{fig:stack}
\end{figure*}

The system under consideration consists of a stack of $M$ fluctuating and interacting membranes of size $L\times L$, as depicted in Fig.~\ref{fig:stack}. The $m$-th membrane's displacement from its average plane is denoted as $u_m(x,y)$, the average distance between membranes $\bar{a}$, and the bending rigidity as $K_c$. Imposing periodic boundary conditions in all directions yields the Hamiltonian of a stack of membranes
\begin{equation}
 \mathcal{H} = \sum_{m = 0}^{M-1} \int \left( \frac{K_c}{2} (\nabla^2 u_m)^2 + \Phi(a_m) \right) dx\,dy,
\end{equation}
where $\Phi$ denotes the bare interaction potential, given here by the hydration repulsion and the van der Waals attraction, and $a_m(x,y) = u_{m+1}(x,y) - u_m(x,y) + \bar{a}$ denotes the local distance between two membranes. We furthermore require $a_m \ge 0$, meaning that membranes cannot interpenetrate. 

To reduce the system's degrees of freedom to a finite amount, the membranes are discretized on a square $N\times N$ lattice. The simulation is performed in the constant pressure ensemble,\cite{mcdonald_npt-ensemble_1972} which converges for this model faster than constant volume simulations \cite{gouliaev_simulations_1998-1}. Monte-Carlo updates are proposed in $\bar{a}$ and in the complex coefficients $u_m(q_x,q_y)$ of the Fourier transformation of $u_m(x,y)$. Simulating in Fourier space allows for larger moves, thereby accelerating equilibration \cite{gouliaev_simulations_1998-1}. After every Monte Carlo step (MCS), which corresponds to degree of freedom ($N^2 M + 1$) update proposals, we re-centered the coordinate system to correct for small center of mass movement as a new feature in the calculations. 

Simulations were performed for $L = \SI{700}{\angstrom}$, several different $N$ (\numlist{6;8;12;16;24;32}), $M = 8$, equilibration lengths of \SI{3e3}{MCS}, and collection lengths of \SI{e4}{MCS}, which typically exceeded the autocorrelation time by a factor of 100. Simulations were started with step sizes estimated from an approximative theory \cite{podgornik_thermal-mechanical_1992} and then subsequently optimized during equilibration, applying either dynamically optimized Monte Carlo (DOMC), or -- as a new feature -- the acceptance ratio method (ARM) as a backup if DOMC fails \cite{gouliaev_monte-carlo_1998, bouzida_efficient_1992}. 

Several observables can be determined from converged simulations, but the two most important quantities for comparison with SAXS experiments are the temporally- and spatially-averaged distance between membranes $d_W = \langle \bar{a} \rangle$ and the time average of its fluctuations 
\begin{equation}
 \Delta^2 = \langle \overline{\left( z_{m+1}(x,y) - z_m(x,y) - d_B - d_W\right)^2} \rangle,
\end{equation}
where the long bar denotes spatial averaging over $(m,x,y)$, $\langle . \rangle$ denotes time averaging, and $z_m(x,y) = u_m(x,y) + m \cdot (\bar{a} + d_B)$ is the $m$-th membrane's position in real-space. 
Specifically, $d_W$ corresponds to the experimental thickness of the water layers separating the lipid bilayers, while $\Delta$ is related to the experimental Caill\'{e} parameter $\eta$ as detailed above. 

It should be emphasized that our explicit purpose of making contact with the X-ray structure factor and the interactions between bilayers, requires much larger systems than can be presently envisioned either for all-atom simulations, used to obtain electron density profiles, or even for the most coarse grained molecular simulations \cite{cooke_solvent-free_2005}.  We require $M$ bilayers in a stack, each bilayer having a large lateral size $L$. It has been shown in previous work \cite{gouliaev_simulations_1998}, that $L = \SI {700}{\angstrom}$ and $M = 8$ are sufficient to obtain accuracies of 1\% for $d_W$ and $\Delta$, and that would require about \num{130000} lipids with associated water in typical molecular simulations. Apart from simulation size, also the necessary timescales, which scale with the fourth power of the undulation wavelength \cite[pp.~77--78]{pabst_liposomes_2014}, render molecular dynamics simulations for that purpose unfeasible. Furthermore, to fit the experimental data requires on the order of 100 separate simulations, distributed on multiple optimizations from different start points. In the membrane MC simulations we employ, each bilayer is reduced to a network consisting of $N$ nodes in each of the two lateral directions and each node has only one degree of freedom.  Computed observables change significantly with $N/L$ \cite{gouliaev_simulations_1998, gouliaev_simulations_1998-1}, so simulations were performed for a sequence of values of $N \in \{6, 8, 12, 16, 24, 32\}$ and then the observables were extrapolated towards $N/L \to \infty$. Further details of this finite size convergence are given in Sec.~S2. 

\subsection{Bare interaction potentials} \label{subsec:bare_potentials}
For uncharged membranes, the potential at bilayer separation $a$ is modeled canonically by \cite{note__additional_steric_interaction}
\begin{equation}
 \Phi(a) \simeq A \lambda \exp \left( -\frac{a}{\lambda} \right) - \frac{H}{12 \pi a^2}. \label{eq:membrane_potential}
\end{equation}
The first term is the well-established empirical form of the solvent-mediated hydration interaction, which has been argued to originate from changes in various measures of order for the water structure at the membrane interface \cite{marcelja_repulsion_1976, kanduc_hydration_2013, kanduc_hydration_2014}, with the strength $A$ and the decay length $\lambda$, which is typically in the range of \SIrange{1}{2}{\angstrom} \cite{petrache_interbilayer_1998}. The second term describes the ubiquitous van der Waals interaction potential for two planar semi-infinite layers, with $H$ being the Hamaker coefficient that in general also depends on the bilayer separation $a$, $H = H(a)$ \cite[p.~15]{parsegian_van_2006}. This functional form is convenient because it can in fact describe both cases of either two finite-thickness layers interacting across a solvent layer \cite{note__van_der_waals_finite_thickness}, as well as effective pairwise interactions in an infinite stack of finite-thickness layers \cite{podgornik_nonadditivity_2006}. For large solvent layer thickness the nonpairwise additive effects in the latter case become negligible and the van der Waals interaction potential for the two cases follows exactly the same separation dependence. 

Due to the divergence of the van der Waals potential for $a \to 0$, the $1/a^2$ term is cut off for $a < \SI{1}{\angstrom}$ \cite{gouliaev_simulations_1998}. In experiments, the collapse of charge neutral bilayers due to van der Waals forces is avoided by very short range steric interactions established by \citeauthor{mcintosh_steric_1987} \cite{mcintosh_steric_1987}, but which occur at significantly higher osmotic pressures than those relevant for the present experiments, see also Fig.~S5. 

To calculate the Hamaker coefficient $H$ {\it ab initio}, we had to approximate the lipid bilayers by pure hydrocarbon \cite{note__van_der_waals_approximations}. Specifically, we calculated $H$ for an infinite stack of hydrocarbon layers in water, based on a full multilayer Lifshitz formulation \cite{podgornik_nonadditivity_2006}. The ranges for the hydrocarbon thicknesses $d_B = \SI{45}{\angstrom}$ to $\SI{60}{\angstrom}$ and the water spacings $d_W = \SI{5}{\angstrom}$ to $\SI{30}{\angstrom}$ were motivated by our experimental data. In this calculation range, differences in the Hamaker coefficient were within 10\%. For our MC simulations the exact value of $H$ matters most when all forces are of comparable magnitude, that is at vanishing external osmotic pressure. We therefore used the $H$ values of $\SI{4.08e-21}{\J} = \SI{4.08}{\zJ}$ for Ld and \SI{4.15}{\zJ} for Lo domains\changed{, see Fig.~\ref{fig:Hamaker}}.

\begin{figure}
  \includegraphics{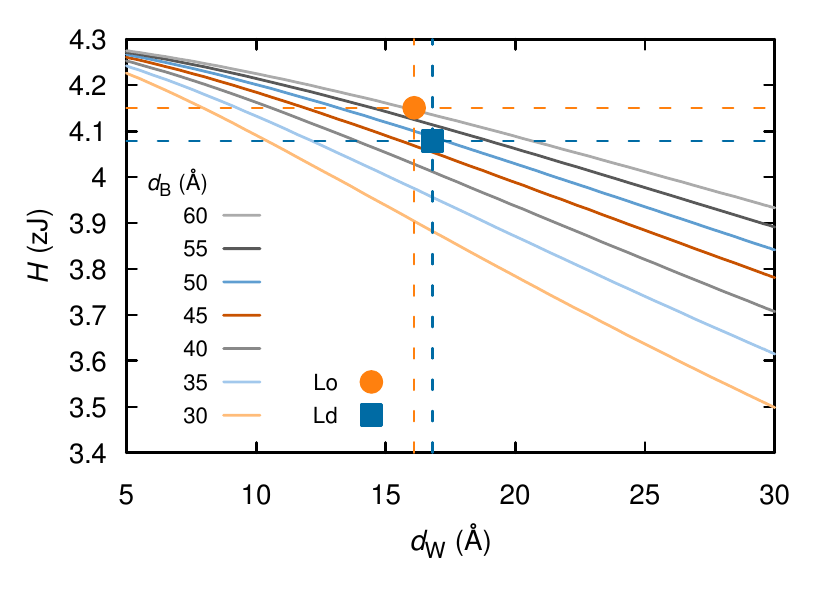}
   \caption{Hamaker coefficient $H$ for hydrocarbon multilayers of height $d_B$ and separation $d_W$ in water. Highlighted are the applied values of $H$ for Ld and Lo, which are described in the main text.}
  \label{fig:Hamaker}
\end{figure}

\changed{
Both components of the bare potential, i.e.\ hydration and van der Waals, cause partial bare pressures between neighboring membranes given by \cite{note__pressure_computation}
\begin{align}
 P_{hyd}(d_W) &= A \exp\left(-\frac{d_W}{\lambda}\right), &
 P_{vdW}(d_W) &= - \frac{H}{6 \pi d_W^3}. \label{eq:partial_pressure_calculation}
\end{align}
For comparison to previous reports using mean-field\slash additivity approximations for modeling undulation interactions, one can obtain an effective decay constant $\lambda_{und}$ by subtracting bare contributions from experimental data, i.e.\ $P_{und} = P - P_{hyd} - P_{vdW}$ \cite{gouliaev_simulations_1998-1}. The undulation decay constant then results from a fit of $P_{und} = A_{und} \, \exp(-\lambda_{und}/d_W)$, with the two adjustable parameters $A_{und}$ and $\lambda_{und}$. Because the undulation pressure deviated significantly from a perfect exponential, we limited the fit to large separations ($d_W \geq$ \SI{14}{\angstrom}). 
}

\subsection{Optimizing parameters against experimental data} \label{subsec:parameter_optimization}
\changed{Calculation of the Hamaker coefficient $H$, as described above, allowed us to reduce the number of free fitting parameters for the simulations to three, $\vec{\Lambda} = (A,\lambda,K_c)$, for a joint analysis of domain separation and fluctuation data (see below).} 

We implemented a least squares routine with Matlab$^{\textrm \textregistered}$ \cite{_matlab_2011}, utilizing its trust region reflective optimization algorithm to minimize the sum of the squared residues 
\begin{equation}
 \chi^2 (\vec{\Lambda}) = \sum_i \left( \frac{d_{W,i} - d_W(P_i; \vec{\Lambda})}{U_\mathrm{eff}(d_{W,i})} \right)^2 + \left( \frac{\Delta_i - \Delta(P_i; \vec{\Lambda})}{U_\mathrm{eff}(\Delta_i)} \right)^2, 
\end{equation}
where $d_{W,i}$ and $\Delta_i$ are the experimentally determined values at fixed osmotic pressure $P_i$, $d_W(P_i; \vec{\Lambda})$ and $\Delta(P_i; \vec{\Lambda})$ are simulation results, and $U_\mathrm{eff}(f)$ is the effective uncertainty of a given quantity $f$, derived from
\begin{equation} \label{eq:effective_uncertainty}
 U^2_\mathrm{eff}(f) = U^2(f_\textrm{exp}) + U^2(f_\textrm{sim}) + \left( \frac{\partial f_\textrm{sim}}{\partial P} \cdot U(P_i) \right)^2.
\end{equation}
The agreement between model and data was evaluated by the reduced $\chi_{red}^2 = \chi^2 / \tilde{N}$, where $\tilde{N}$ equals the number of data points minus the number of free parameters \cite[p.~268]{taylor_introduction_1997}. The Jacobian for this gradient based algorithm and the derivative in Eq.~\eqref{eq:effective_uncertainty} were computed with the histogram reweighting method described in Sec.~S3. Once the iteration converged, the uncertainties of the fit parameters were determined from the curvature of $\chi_{red}^2$. In order to locate the global optimum, several iterations from randomly chosen initial parameter sets were performed. 

To test our implementation, we fitted simulation results determined for one reasonable parameter set $\vec{\Lambda}'$, by starting the least squares from several different initial starting points $\vec{\Lambda}$. Within 3--5 iterations, these optimizations converged towards the correct values $\vec{\Lambda}'$, thereby indicating that the weighted histogram based differentiation and the fit were correctly implemented. For the experimental data sets, convergence was usually reached within 10 iterations. However, due to the stochastic nature of the simulations and the consequential randomness of results and derivatives, the optimization algorithm propagated poorly in flat regions, i.e.\ small $\vec{\nabla} \chi_{red}^2$. Because $\chi_{red}^2(\vec{\Lambda})$ is a smooth function and its gradient has to vanish at extrema, the optimization algorithm's efficiency decreased, the closer it got to the optimum. This was another reason for starting several independent iterations \cite{note__optimization_for_simulation}. 

As a further test case, we re-analyzed previously published osmotic pressure data of pure dimyristoyl-phosphocholine (DMPC) bilayers \cite{petrache_interbilayer_1998}, yielding very reasonable values and a good agreement between simulations and experiments. Details are given in Sec.~S4. \changed{Thus, we conclude that our method provides a robust analysis for interactions in fluctuating membrane assemblies.}



\section{Results and Discussion}
\subsection{X-ray analysis} \label{subsec:x-ray_analysis}
SAXS patterns were analyzed as detailed previously \changed{by a Caill\'{e} theory-based analysis  \cite{heftberger_situ_2015}.} Figure~\ref{fig:saxs} showcases the analysis for two samples at osmotic pressures of \SI{34}{\kPa} and \SI{2.4}{\MPa}, demonstrating that shapes and positions of Bragg reflections are well reproduced. 
\changed{Consistent with previous studies \cite{pabst_effect_2009, yuan_solution_2009, heftberger_situ_2015}, we find sharper and more prominent Bragg reflections for the Lo phase due to its decreased bending fluctuations, compared to the coexisting Ld phase.}
Fits for all other samples are shown in Sec.~S5. For increased \changed{osmotic} pressures, Bragg peaks shifted towards higher $q$ and became more prominent. This is due to the decrease of bilayer separation which goes in hand with a reduction of bending fluctuations in agreement with previous reports \cite{petrache_interbilayer_1998, hemmerle_controlling_2012}. 

\changed{Peak line-shapes for Lo and Ld domains were found to be well described by the applied Caill\'{e} theory, particular at low osmotic pressure (Fig.~S4). Since this theory is incapable of fitting peaks from lamellar gel phases \cite{pabst_global_2006}, we conclude that neither peaks assigned to the Lo, nor to the Ld phase can originate from a gel phase.
This is also consistent with reported compositional DSPC/DOPC/Chol phase diagrams \cite{zhao_phase_2007, heberle_comparison_2010} and a recent SAXS study from our lab, reporting for the identical lipid mixture that the structural parameters match those of Lo and Ld phases at the tie-line endpoints \cite{heftberger_situ_2015}. 
}

\changed{Fit quality of SAXS spectra worsened for increased PEG concentrations, see Fig.~\ref{fig:saxs} or Sec.~S5. Probably the underlying Caill\'{e} theory loses its applicability for the increased order experienced at elevated osmotic pressures. While effects on domain separation were negligible, fluctuations determined from the fits became increasingly skewed with osmotic pressure, in particular for Lo domains (see below). }

\begin{figure}
  \includegraphics{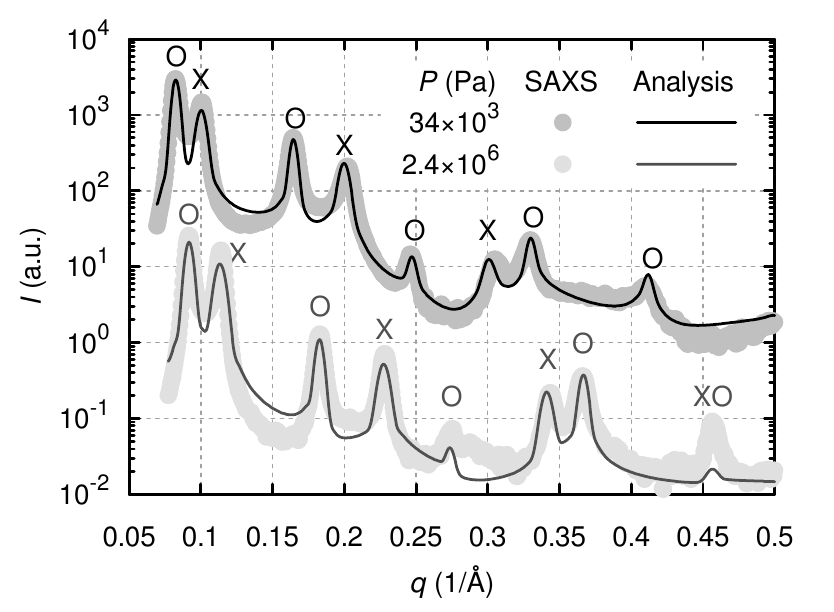}
  \caption{Calculated scattering intensities (solid lines) from full $q$-range analyses, compared with recorded SAXS data from coexisting phases (dots) for two different osmotic pressures $P$. Bragg reflections from aligned Lo and Ld domains are indicated by symbols O and X, respectively.}
  \label{fig:saxs}
\end{figure}

The effect of osmotic pressure on the lamellar repeat spacing $d$, as determined from the SAXS analysis, is plotted in Fig.~\ref{fig:PvsD_Ld_Lo}. At high osmotic stress, the distance between bilayers is effectively set by the repulsive hydration interaction which dominates the repulsive fluctuation interaction and the attractive van der Waals interaction. As osmotic pressure is decreased, the water spacing between bilayers $d_W$ increases and the fluctuation interaction eventually dominates the hydration interaction. As the osmotic pressure is reduced to zero, the attractive van der Waals force balances the total repulsive forces, resulting in finite $d_W$ and $d$ values.

\begin{figure}
  \includegraphics{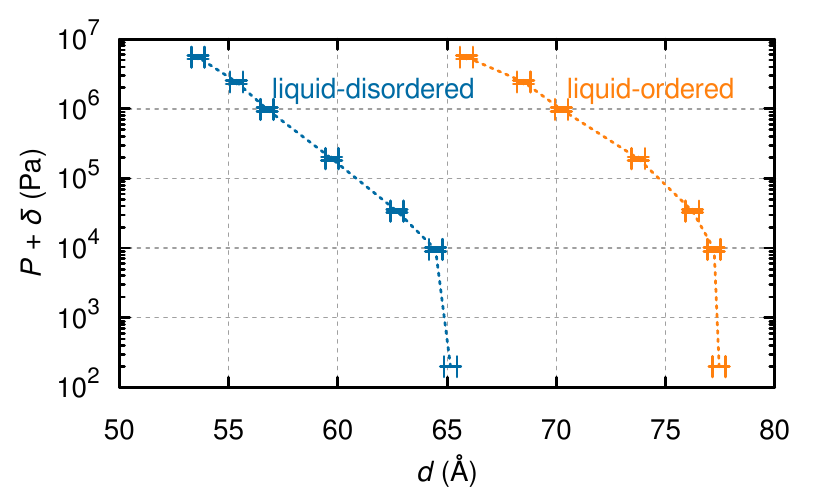}
  \caption{Osmotic pressure $P$ vs membrane periodicity $d$ for Ld and Lo determined by SAXS analysis \cite{note__small_offset}. Dashed lines are meant solely as a guide for the eye.}
  \label{fig:PvsD_Ld_Lo}
\end{figure}

Within experimental uncertainty, the two isotherms in Fig.~\ref{fig:PvsD_Ld_Lo} are rather similar when the difference in membrane thickness is taken into account ($d_B^{Ld} = \SI{48.5(10)}{\angstrom}$ and $d_B^{Lo}=\SI{61.3(12)}{\angstrom}$). Of course, identical isotherms would imply that all the interactions are identical.  However, significant experimental differences were observed in the fluctuation behavior as detailed below, corroborating the crucial advantage of jointly analyzing fluctuations and osmotic pressure isotherms in order to obtain the interaction parameters \cite{petrache_interbilayer_1998}.

\subsection{Optimized simulations}
The experimental data and the results of optimized simulations are compared in Fig.~\ref{fig:mmc-saxs2}, while Tab.~\ref{tbl:mmc_fit_results} lists \changed{results for the interaction parameters}. Experimental errors for $d_W$ and $\eta$ were obtained from the SAXS analysis and for $P$ were estimated to equal the pipetting error of 6\% for viscous PEG solutions. To quantify the agreement between data and simulations, we report $\chi_{red}^2$, which becomes ca.\ 1 if the differences are compatible with experimental errors \cite[p.~268]{taylor_introduction_1997}. This is the case for the Ld phase, where simulations and experimental data match ideally, but the mismatch for Lo is bigger than expected ($\chi_{red}^2 = 6$). 

We are inclined to attribute this discrepancy for Lo at least partially to the limited applicability of the Caill\'{e} theory for highly ordered systems, as described in the previous section. Indeed, deviations in $\Delta$ are especially pronounced for small bilayer separations, i.e.\ at high osmotic pressures. In light of these discrepancies, we suggest that the experimental uncertainties determined for the Lo phase are rather too small because they do not take into account the decreasing applicability of the Caill\'{e} theory for more ordered phases whose fluctuations are suppressed by low hydration.

\begin{figure*}
  \includegraphics[width=\textwidth]{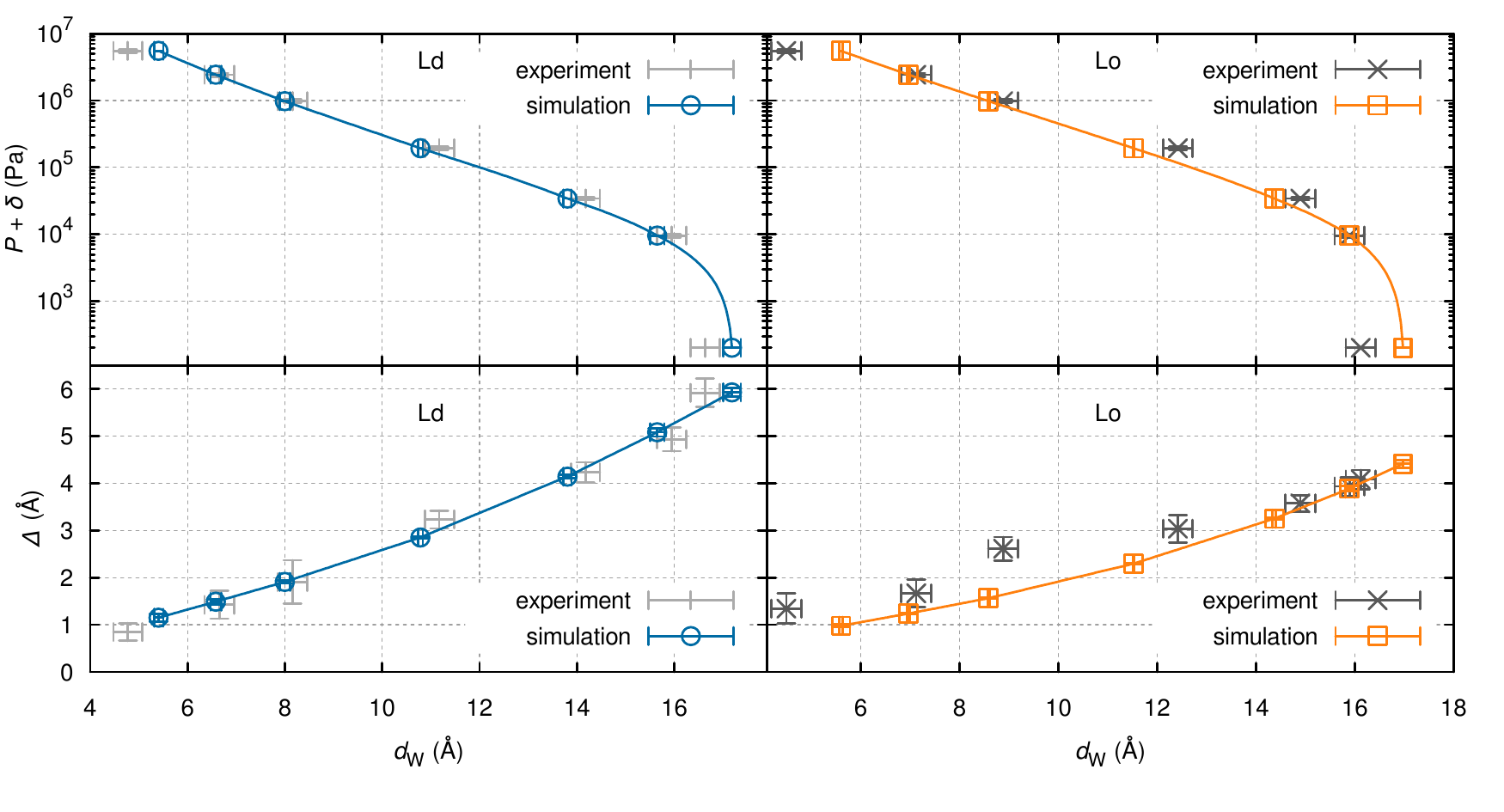}
  \caption{Osmotic pressure (top) and fluctuations (bottom) vs water-layer thickness for best fit of membrane MC simulation (cyan/orange) against SAXS data (gray) \cite{note__small_offset}. Solid lines were obtained by exponentially interpolating fluctuation contributions.}
  \label{fig:mmc-saxs2}
\end{figure*}

\begin{table}
  \caption{Optimal parameters determined for describing the coexisting Lo/Ld phases in DOPC/DSPC/Chol (0.42:0.37:0.21). Errors as obtained from the fitting routine, see text for further details.}
  \label{tbl:mmc_fit_results}
  \begin{tabular}{l ll}
    \hline
     & Ld & Lo \\
    \hline
    $K_c / \si{\zJ}$ & \num{44(10)} & \num{120(20)} \\
    $A / \si{\Pa}$ & $10^{\num{8.3(2)}}$ & $10^{\num{8.1(2)}}$ \\
    $\lambda / \si{\angstrom}$ & \num{1.37(15)} & \num{1.74(15)} \\
    $\chi_{red}^2$ & \num{1.5(5)} & \num{5.8(5)} \\
    \hline
  \end{tabular}
\end{table}

While differences in $P(d_W)$ are insignificant between Ld and Lo (see also Fig.~\ref{fig:PvsD_Ld_Lo}), fluctuations of the Lo phase, containing most of the DSPC and about thrice as much cholesterol as Ld, are evidently smaller (Fig.~\ref{fig:mmc-saxs2}). In the continuum mechanics treatment used in the simulations, this increase in bilayer stiffness is captured by a threefold higher $K_c$ for Lo, see Tab.~\ref{tbl:mmc_fit_results}. 

The values obtained by us for $K_c$ compare well with previously reported results from different techniques. Bending rigidities of binary DOPC/cholesterol mixtures have been measured by several groups, ranging from (\num{60(8)}) to \SI{100(25)}{\zJ} and were found to be largely unchanged by the cholesterol content \cite{pan_cholesterol_2008, sorre_curvature-driven_2009, tian_bending_2009, gracia_effect_2010}. This supports the $K_c = \SI{44(10)}{\zJ}$ obtained for Ld, where DOPC is the main constitutent \cite{heberle_comparison_2010}.  In contrast, a larger concentration of saturated lipids, for which $K_c$ does increase with cholesterol \cite{pan_cholesterol_2008}, is present in the Lo phase, so a larger bending rigidity would be expected for Lo than for Ld. Our finding of $K_c$ = \SI{120(20)}{\zJ} for the Lo phase is consistent with this expectation.

Furthermore, molecular dynamics (MD) simulation results are available for comparison. Khelashvili et al.  \cite{khelashvili_calculating_2013} used the reported tie-line endpoint compositions \cite{heberle_comparison_2010} to separately simulate the liquid-disordered and -ordered phases, obtaining bending moduli of \SIrange{80}{130}{\zJ} for Ld and \SIrange{340}{440}{\zJ} for Lo. Although these values are large compared to our results, both methods find a strong increase of $K_c$ between Ld and Lo.

In agreement with Ref.~\citenum{pan_effect_2009}, we find that a rather simple model suffices to relate bending to area extension moduli for cholesterol-rich samples \cite{note__evans_suggestion}. Based on the assumption that the main contribution to membrane rigidity comes from the stiff cholesterol ring of size $\delta'$, \citeauthor{pan_effect_2009} used the relationship $\delta'^2 = 12 K_c / K_A$. For our samples, with $K_A$ = \SIlist{430;2100}{\mN \per \m} (see Sec.~S1 for details), this equation yields $\delta'$ = \SIlist{11;8}{\angstrom} for Ld and Lo, respectively, in good agreement to actual cholesterol ring sizes of about \SI{9}{\angstrom}, giving additional support to our analysis.  

\subsection{Interdomain forces}
As stated before, the differences between Ld and Lo in the $P$ vs $d_W$ data sets are small. However, a more thorough investigation of these quantities yields interesting insights. Because good fits to these data were obtained, the total pressure $P$ is readily dissected into its individual contributions from the fundamental surface forces\changed{, whose functional dependences are} plotted in Fig.~\ref{fig:force-dissection}.

\begin{figure}
  \includegraphics{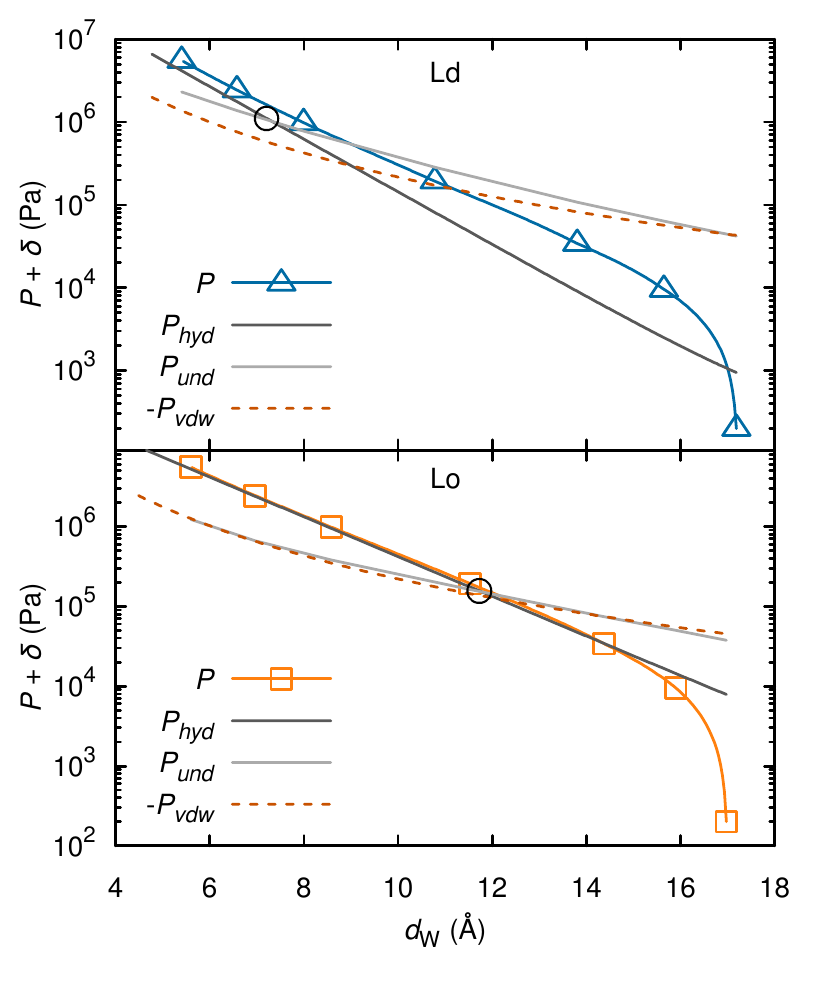}
  \caption{Partitioning of total pressure $P$ into contributions from hydration $P_{hyd}$, van der Waals $P_{vdw}$, and undulations $P_{und}$ for Ld (top) and Lo (bottom) \cite{note__small_offset}. The large open black circles show the values of the separation $d_W$, at which hydration and undulation pressure are equal. Due to the additive constant $\delta$, the hydration pressure deviates from a straight line at low $P$.}
  \label{fig:force-dissection}
\end{figure}

The thicker Lo bilayer causes an increase in the Hamaker coefficient, but only by 3\% compared to the Ld phase; this is a minor difference in the van der Waals interaction that is hardly noticeable in the $P_{vdW}$ curve in Fig.~\ref{fig:force-dissection}. For small bilayer separations, the hydration interactions are of similar magnitude and represent, as expected, the dominant contribution to the total interaction potential for both phases. Despite these similarities, the fluctuation pressure starts to surpass the hydration pressure already at much smaller separations $d_W$ for Ld than for Lo. This difference implies, in contrast to the ordered phase, that the undulation interaction becomes the most important repulsive interaction over a wider range of bilayer separations in the case of the disordered phase. Stronger repulsions due to fluctuation interactions are of course reasonable because thermal undulations were found to be significantly increased for the Ld phase (Fig.~\ref{fig:mmc-saxs2}). Nevertheless, even in the Lo phase, the thermal undulation interaction dominates the hydration force over the most important, well hydrated range of $d_W$, starting at separations of \SI{12}{\angstrom}. 

We obtained almost exponentially decreasing fluctuation forces of the scaling form $\propto \exp{(-z/\lambda_{und})}$, with effective decay lengths $\lambda_{und} \approx$ \SIlist{3.3;3.7}{\angstrom} for Ld and Lo, respectively The ratio of fluctuation to hydration decay length $\lambda_{und}/\lambda$ is obtained as 2.4 for Ld and 2.1 for Lo. While the mean-field theory predicted its value as 2.0 \cite{podgornik_thermal-mechanical_1992}, values of 2.4 have been reported for simulations \cite{gouliaev_simulations_1998-1, gouliaev_simulations_1998}, and \numrange{2.0}{3.0} from other experiments \cite{petrache_interbilayer_1998, pabst_entropy-driven_2007, pabst_effect_2009}. 

Compared to Lo, a significantly shorter decay length for the hydration interaction pressure was found for the Ld phase. At present, the origin for this difference is unclear. However, it is this difference combined with the larger fluctuation force that gives $P$ versus $d_W$ curves that are nearly the same for Lo and Ld, both with fully hydrated $d_W$ close to \SI{17}{\angstrom}.

Domain alignment across interlamellar aqueous phases has recently been hypothesized to be caused by water network mismatch due to the different hydration properties of Lo and Ld phases \cite{tayebi_long-range_2012}. In support of this postulation, we observed significantly different hydration forces and nearly equal van der Waals forces for both phases. Thermal fluctuations were however neglected in the aforementioned hypothesis, while we now find considerable differences specifically in the undulation forces for coexisting domains. Their importance is especially striking near full hydration, where undulation and van der Waals pressures surpass hydration repulsion by an order of magnitude (see Fig.~\ref{fig:force-dissection}).



\section{Conclusion}
We have evaluated the fundamental long-range interactions between \changed{coexisting} Lo and in Ld domains in DOPC/DSPC/cholesterol\changed{, which is a frequently used model system for mammalian outer plasma membranes \cite{veatch_separation_2003, scherfeld_lipid_2003, zhao_phase_2007, heberle_comparison_2010, heberle_bilayer_2013, heftberger_situ_2015, heberle_phase_2011, marsh_cholesterol-induced_2009}}.  Because we could do this at concentrations where Lo and Ld domains coexist, we were able to avoid all uncertainties in the phase diagram and its associated tie-lines between Lo and Ld phases. This work combines methodology  from three separate inputs: SAXS\slash osmotic stress experiments, comprehensive Monte Carlo simulations, and detailed calculations of van der Waals interactions. 

The reported values for fundamental surface forces and bending moduli are the first of their kind being, directly obtained from coexisting Lo/Ld domains. The underlying full $q$-range SAXS analysis allowed us to quantify the extent of fluctuations and capture their dependence on osmotic pressure, which proved essential for determining the bending rigidities of cholesterol-rich phases. We obtained bending moduli of \SI{44}{\zJ} for Ld and a roughly threefold higher value for Lo domains, attributable to their larger concentrations of saturated lipid and cholesterol. 

While we obtained almost identical van der Waals interactions for aligned Lo and Ld domains, the remaining interactions, however, turned out to be strikingly different: decay lengths of the hydration pressures differed by 25\% between Lo and Ld phases, and repulsions due to thermal fluctuations were found to be significantly increased for Ld. \changed{These findings provide evidence that a combination of hydration repulsion and the fluctuation-driven undulation repulsion must be considered in any quantitative explanation of the long-range positional correlations between aligned Lo and Ld domains. In particular the strong entropic contribution from undulating Ld domains may be a leading term to be considered. We therefore expect that our study will form the base for a concise theory of domain alignment.} 


\section*{Author Contributions}
B.K. designed and performed research, analyzed data and wrote the paper; P.H. designed and performed research and analyzed data; R.P. and J.F.N. contributed analytic tools and wrote the paper; G.P. designed and performed research and wrote the paper.

\section*{Acknowledgments}
This work is supported by the Austrian Science Fund FWF, Project no.~P24459-B20 to GP. The computational results presented have been achieved using the Vienna Scientific Cluster (VSC). The authors thank Alexander Rieder and Heinz Amenitsch for experimental support and Hans Gerd Evertz for critical review of the simulation and advice regarding finite size convergence. Support for the original development of the MC software was provided to JFN under grant GM44976 from the U.S.\ National Institutes of Health. RP would like to acknowledge the SLO-A bilateral grant N1--0019 of the Slovene Research Agency. 

\section*{Supporting Citations}
\nocite{podgornik_solvent_1988, ben-tal_robust_1999, ben-tal_robust_2000, fu_feature_2002, ben-tal_adjustable_2004} 
References \cite{salsburg_application_1959, ferrenberg_new_1988, ferrenberg_optimized_1989, kumar_weighted_1992, nagle_introductory_2013, chu_anomalous_2005, bonner_linear_1964, nagle_numerical_1970, quenouille_notes_1956, tukey_bias_1958} appear in the Supporting Material. 


\begin{thebibliography}{103}
\bibitem[not({\natexlab{a}})]{note__supplier_provided_M_W}
Supplier provided $M_W$, accounting for an additional water molecule with DOPC,
  were used for determining stock concentrations.
\bibitem[not({\natexlab{g}})]{note__pressure_computation}
\changed{Equation \eqref{eq:partial_pressure_calculation} was derived from
  $P_j(d_W) \approx -\partial \Phi_j(d_W)/\partial d_W$. The difference to the
  exact relationship $P_j(d_W) = \left< -\partial \Phi(\bar{a})/\partial
  \bar{a} \right>$ was found to be less than the simulational uncertainty.}
\bibitem[not({\natexlab{i}})]{note__small_offset}
The small offset $\delta = \SI{200}{\Pa}$ is necessary for plotting $P = 0$ on
  a logarithmic scale.
\end{thebibliography}

\clearpage




\section*{Supplementary material}

# Supporting information for: Bending rigidities and interdomain forces in membranes with coexisting lipid domains

B. Kollmitzer, P. Heftberger, R. Podgornik,
J. F. Nagle, and G. Pabst

April 15, 2015, Graz

## S1 Area extension modulus estimation

The dependence of bilayer thickness on osmotic pressure $P$ is accounted for via the area extension modulus $K_A$ and given by the equation[1]

$$d_B(P) = d_B(0) \frac{K_A + P\, d(P)}{K_A + P\, d_B(0)}. \tag{S1}$$

We estimated $K_A$ for our coexisting liquid phases based on published data for single lipids and binary lipid mixtures by Rawicz *et al.*[2,3] The Ld phase under investigation consists essentially of DOPC, with approximately $10\,\text{mol}\%$ cholesterol.[4] Interpolating linearly between the two published values for 0 and $50\,\text{mol}\%$ cholesterol in DOPC[2] yields $K_A(\text{Ld}) = (430 \pm 30)\,\text{mN}\,\text{m}^{-1}$.

In the coexisting Lo phase, the main constituent is the saturated lipid DSPC, which is accompanied by ca. $30\,\text{mol}\%$ cholesterol.[4] Unfortunately, published $K_A$ values for saturated lipids are sparse. As a compromise, we interpolated linearly between pure DMPC ($0\,\text{mol}\%$ cholesterol) and a 1:1 mixture of sphingomyelin/cholesterol,[2,3] yielding $K_A = (2100 \pm 500)\,\text{mN}\,\text{m}^{-1}$ for our Lo phase.

As pointed out in the section *X-ray measurements* of the main text, knowing the magnitude of $K_A$ is more important than getting the precise number. That is because the biggest estimated change in bilayer thickness turned out to be just $0.3\,\text{Å}$. In principle, such a subtle difference in $d_B$ would be resolvable with SAXS, but not with the additional scattering signal due to PEG.

## S2 Finite size convergence

With open edges, one generally expects a 'surface' perturbation proportional to the relative size of the boundary to the interior, i.e. proportional to $1/N$ for our systems. As is

well known, periodic boundary conditions generally reduce this perturbation. They also speed up the convergence with system size, from $1/N$ to $1/N^2$ in a case well documented by Bonner and Fisher[5] (note their Fig. 1) and in the case of the one-dimensional Ising model the convergence is exponentially fast with periodic boundary condions. While another case with very slow convergence is known,[6] that one is due to very long range interactions not present in our membrane stacks. For periodic boundary conditions, the exact solution of a harmonic approximation to Eq. (3) suggests that $d_W$ and $\Delta$ converge asymptotically like $y(N) \sim c_\infty - c_2/N^2$, i.e. convergence is expected to be faster than $1/N$ and, in agreement with the previous simulations,[7] our results are consistent with a dominant $1/N^2$ asymptotic convergence, allowing, of course, for higher order terms.

We perform simulations for several 'densities' $N \in \{N_{min}, \ldots, N_{max}\}$ and fit them with the function $y(N) = c_\infty + \sum_{k=2}^{k_{max}} c_k/N^k$. Together with the originally proposed $k_{max} = 3$ and $N \in \{6, \ldots, 32\}$, this method yields sufficiently precise continuum estimators $c_\infty$, compared to the experimental uncertainties.[8] However, we found that varying the arbitrary parameters $k_{max}$ and $N_{min}$ influenced the final estimator stronger for some simulations (e.g. high pressures) than for others. To obtain more reliable uncertainties and perhaps even better continuum estimates, we perform now several extrapolations, with different values for $k_{max}$ and $N_{min}$, but always using the highest possible $N_{max}$. By not changing $N_{max}$, we weight the most significant simulations (with the highest density) stronger. This procedure yields a list of results for $c_{\infty,l}$, which we average for the final estimator. Its uncertainty is then determined by the individual errors of $c_{\infty,l}$ (statistical uncertainty of observables due to finite simulation length) and their standard deviation (error due to finite simulation density). This procedure is closely related to the Jackknife technique.[9,10]

Comparisons between these improved Jackknife estimators and estimators obtained by the original method are given in Fig. S1. The relative difference in the estimators were less than 5% for all performed simulations, but most importantly, Jackknife produces a meaningful uncertainty.

## S3 Efficient differentiation

A single simulation of a particular set of parameters $\vec{\Lambda} = (P, A, H, \lambda, K_c, \ldots)$ contains more information in the generated time series, than the aforementioned observables which are determined by averaging. By reweighting the simulated histogram of density of states, it is possible to compute these quantities over a certain range of simulation parameters and thereby also derive their gradients.[11–13][14] This well recognized method was briefly mentioned for membrane MC simulations,[15] but has not been implemented for them previously.

We calculated the expectation value of an observable $f(u, \bar{a})$ for a different set of parameters $\vec{\Lambda}'$ from a simulation performed at $\vec{\Lambda}$ by

$$\langle f \rangle_{\vec{\Lambda}'} = \frac{\sum f_{\vec{\Lambda}'}(u, \bar{a}) \cdot \exp\left(-\delta G/kT\right)}{\sum \exp\left(-\delta G/kT\right)}, \tag{S2}$$

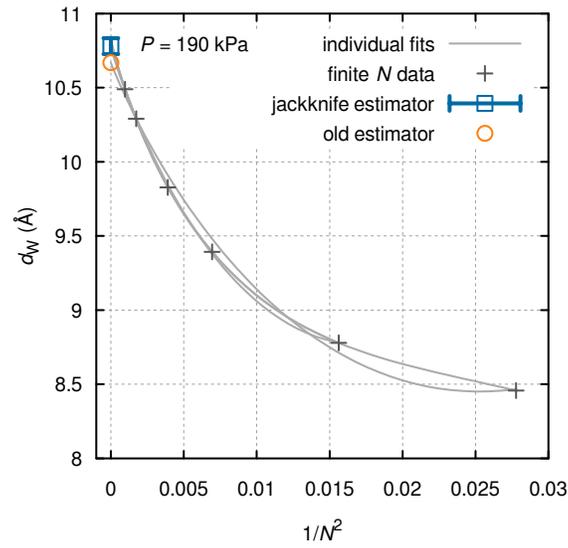

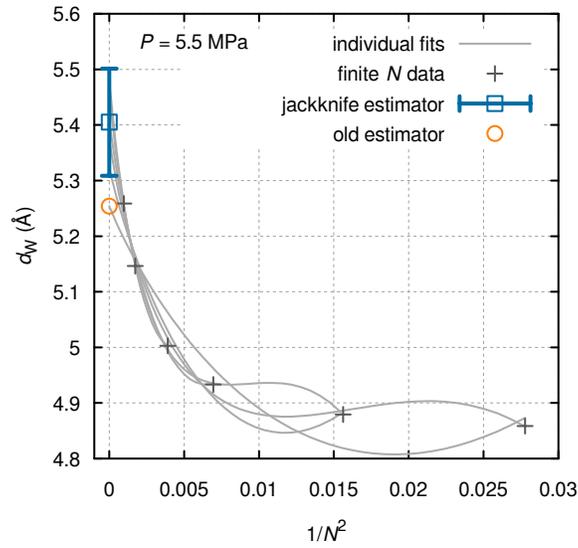

Figure S1: Finite size convergence of membrane spacing $d_W$ vs membrane "density" $N$ of Ld domains according to Tab. 1 at intermediate (top) and high osmotic pressures $P$ (bottom). A variant of Jackknife allows us to obtain reasonable errors for the estimator. Statistical uncertainties for plotted finite $N$ data are less than $10^{-2}\,\text{Å}$.

where the sums extend over all realized configurations and $\delta G$ is the change in the Gibbs energy of each state $(u, \bar{a})$ upon changing $\vec{\Lambda}$ to $\vec{\Lambda}'$. Most parameters could be separated from $u$ and $\bar{a}$ in our case, yielding $\delta G(u, \bar{a}) = \delta \Lambda \cdot \xi(u, \bar{a})$. This allowed us to store only the time series of $\xi$ instead of all realized states. The parameters $P$, $A$, $H$, and $K_c$ were separable in this way, yielding

$$\frac{\delta G}{V} = \delta P \, \xi_P + \delta A \, \lambda \, \xi_A - \frac{\delta H}{12\pi} \, \xi_H + N^2 \frac{\delta K_c}{2} \, \xi_{Kc}, \tag{S3}$$

where $\xi_P = \bar{a}/d_W$, $\xi_A = \overline{\exp(-a/\lambda)}$, $\xi_H = \overline{1/a^2}$, and $\xi_{Kc} = \overline{q^4 |u_m(q_x, q_y)|^2}$. The local distance between membranes is denoted by $a = u_{m+1}(x, y) - u_m(x, y) + \bar{a}$, while the bars denote averages over $(m, x, y)$ or $(m, q_x, q_y)$. $V = L^2 M \bar{a}$ is the membrane stack's volume.

Separating $\lambda$ from $(u, \bar{a})$ in $\delta G$ turned out to be impossible, but we were able to calculate gradients of $d_W$ and $\Delta$ with respect to $\lambda$ efficiently. Because $d_W$ and $\Delta$ did not depend explicitly on $\lambda$ (i.e. $\partial f / \partial \lambda = 0$), differentiating Eq. (S2) yielded,

$$\left. \frac{\partial \langle f \rangle_{\lambda'}}{\partial \lambda'} \right|_{\lambda' = \lambda} = -\frac{AV}{kTN^2\Omega} \left( \sum f(u, \bar{a}) \xi_\lambda - \langle f \rangle_\lambda \sum \xi_\lambda \right), \tag{S4}$$

where sums extend over all realized states, $\Omega$ denotes the collection length and

$$\xi_\lambda = \overline{\left(1 + \frac{a}{\lambda}\right) \exp\left(-\frac{a}{\lambda}\right)}. \tag{S5}$$

Up to first order, $\langle f \rangle_{\lambda'}$ was then determined from $\langle f \rangle_{\lambda'} \approx \langle f \rangle_\lambda + (\lambda' - \lambda) \, \partial \langle f \rangle / \partial \lambda$.

Thus, for any observable $f \in \{d_W, \Delta\}$ and parameter $\Lambda \in \{P, A, H, K_c, \lambda\}$, we first determined $\langle f \rangle_{1,2}(N)$ for $\Lambda_{1,2} = \Lambda \pm \delta \Lambda$ as detailed above, extrapolated these expectation values for $N \to \infty$ according to section S2, and finally calculated the finite difference quotient $\partial \langle f \rangle / \partial \Lambda \approx (\langle f \rangle_1 - \langle f \rangle_2)/2\delta\Lambda$. Relative finite differences were set to $\delta\Lambda/\Lambda = 0.03$.

We checked this method against direct numerical differentiation for a couple of reasonable parameters. Errors were always sufficiently small (well below 50%) to lead the optimization routine towards a global minimum (see the section *Optimizing parameters against experimental data* of the main text).

## S4 Results for a homogeneous control sample

We tested our analysis on already published SAXS data for homogeneous DMPC MLVs determined at $30\,°\mathrm{C}$.[16] The Lifshitz calculation of the van der Waals forces yielded a value of $H = 4.11\,\mathrm{zJ}$ for the published bilayer thickness of $44.0\,\text{Å}$. The obtained values describing the intersurface forces are given in Tab. S1, while Fig. S2 compares the simulations with the experimental data. Reassuringly, the simulations fit the experimental osmotic pressure data well. While the fit to $\Delta$ is excellent for high hydration, the fit becomes relatively poor for $\Delta$ as $d_W$ becomes small, similarly to our $L_o$ sample and likely for the same reason given in the main text.

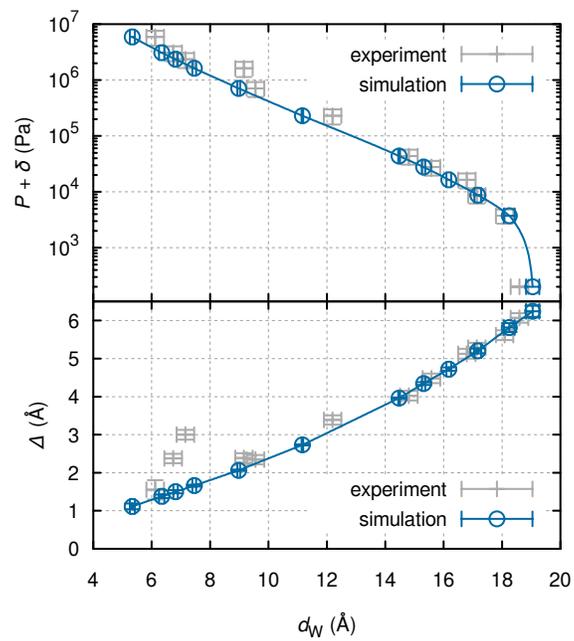

Figure S2: Osmotic pressure (top) and fluctuations (bottom) vs water-layer thickness for best fit of membrane MC simulation (cyan) against SAXS data (light gray) obtained from Ref. 16.[17] Solid lines were obtained by exponentially interpolating fluctuation contributions.

The interaction parameters obtained from the fit are shown in Tab. S1. Literature values for DMPC's bending modulus range from 50–130 zJ at 30 °C.[18] In light of this large variation, comparing only results of related methods is appropriate. Ref. 16 could not determine $K_c$ and the modulus $B$ separately and therefore considered several values of $K_c$; two of these are shown in Tab. S1. The values of $A$ agree very well with ours. The larger values of $\lambda$ would have been smaller if the true value of $K_A$ had been known at that time. Two differences from the previous analyses are that here we calculated $H$ and we used simulations; these cause the main differences reflected in the pairs of values for $H$ and $K_c$ in Tab. S1. Table S1 also shows results from another study,[19] that employed the same kind of simulations used here and differed by obtaining X-ray data from oriented stacks of DMPC bilayers, from which $K_c$ was obtained directly. It also used the same $P$ data, but failed to readjust the $A$ and $\lambda$ values to account for the corrected $K_A$. Nevertheless, agreement is reasonable.

Table S1: Optimal parameters found for describing the DMPC data published in Ref. 16.

|  | Current | 1998a[16] | 1998b[16] | 2005[19] |
|---|---|---|---|---|
| $H$/zJ | 4.11 | 7.13 | 4.91 | 6.1 |
| $K_c$/zJ | $57 \pm 5$ | 50 | 80 | 69 |
| $A$/Pa | $10^{8.1 \pm 0.2}$ | $10^{8.1}$ | $10^{8.1}$ | $10^{8.1}$ |
| $\lambda$/Å | $1.66 \pm 0.15$ | 1.91 | 1.97 | 1.91 |

For completeness, the functional dependence of the individual fundamental surface forces for DMPC is plotted in Fig. S3. The fluctuation force becomes the dominant repulsive force when $d_W$ exceeds 9 Å, intermediate between the values of the Ld and Lo phases in Fig. 7, suggesting that the DMPC bilayer fluctuations are intermediate in this regard between the more fluid Ld phase and the more ordered Lo phase in the studied mixture. This is consistent with the Ld phase having a high concentration of the more disordered unsaturated lipids and the Lo phase having longer saturated chains with cholesterol.

## S5 SAXS analysis

Comparisons between full $q$-range SAXS analyses and experimental data are shown in Fig. S4. Deviations between data and fits, especially for higher $q$ ranges, are due to imperfect background subtraction, as explained in the section *X-ray measurements* in the main text.

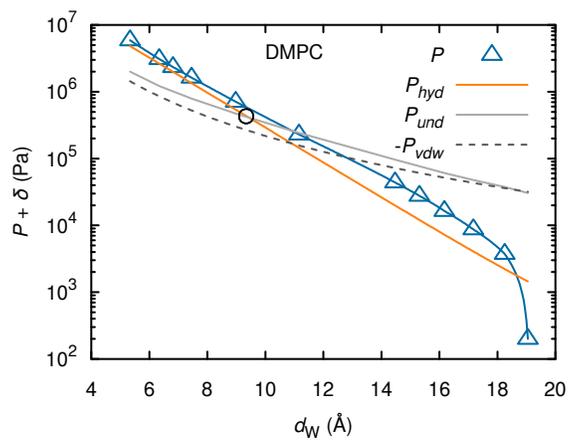

Figure S3: Partitioning of total pressure $P$ into contributions from hydration $P_{hyd}$, van der Waals $P_{vdw}$, and undulations $P_{und}$ for DMPC.[17] The large open black circle shows the value of the separation $d_W$ at which hydration and undulation pressure are equal.

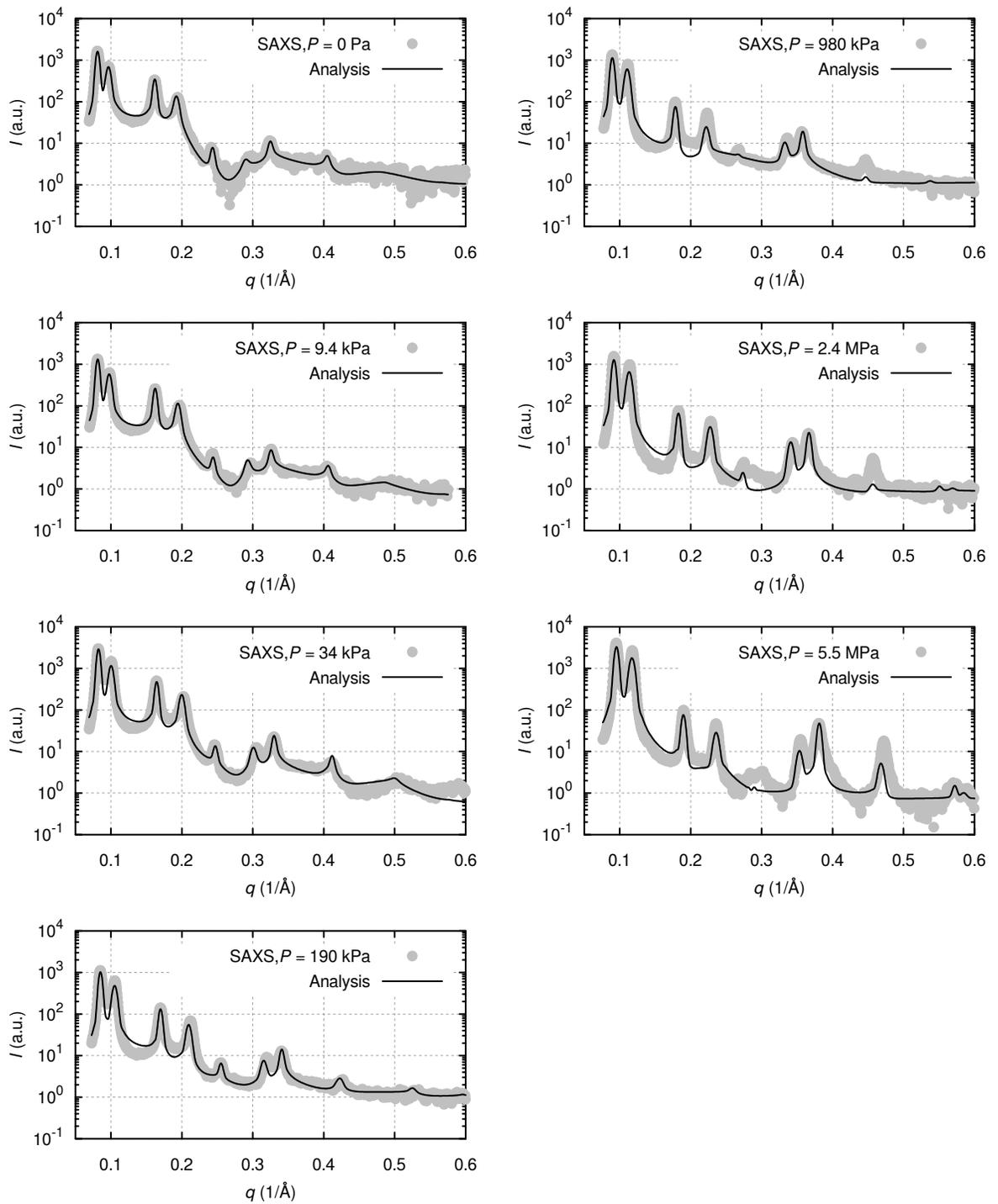

Figure S4: Calculated scattering intensities (solid lines) from full $q$-range analyses, compared to recorded SAXS data from coexisting phases (dots) of DOPC/DSPC/Chol (0.42:0.37:0.21), for all recorded osmotic pressures $P$.

# S6 Fluctuations of the interbilayer water spacing

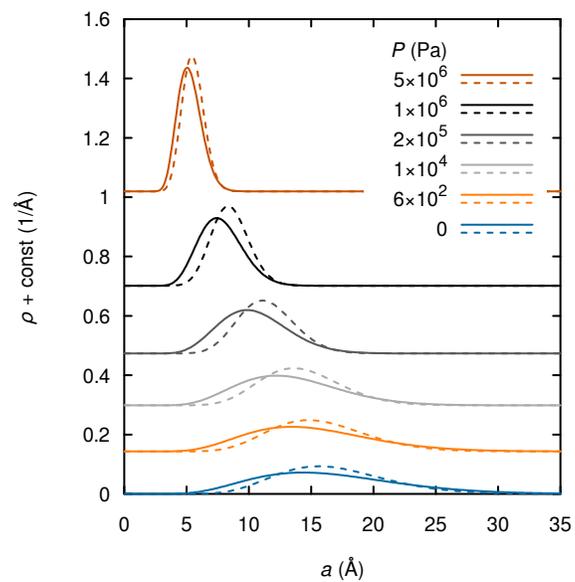

Figure S5: Probability density function $\rho$ of the water spacing $a$ at different external pressures $P$, for Ld (solid) and Lo (dashed) according to Tab. 1, obtained from $N = 32$ simulations.


\end{document}